\documentclass[nonacm, sigconf]{acmart} 
\renewcommand\footnotetextcopyrightpermission[1]{} 
\AtBeginDocument{%
  \providecommand\BibTeX{{%
    \normalfont B\kern-0.5em{\scshape i\kern-0.25em b}\kern-0.8em\TeX}}}

\setcopyright{acmcopyright}
\copyrightyear{2020}
\acmYear{2020}
\acmDOI{10.1145/1122445.1122456}

\acmConference[CIKM 2020]{The 29th ACM International Conference on Information and Knowledge Management}{October 19--23, 2020}{Galway, Ireland}
\acmBooktitle{The 29th ACM International Conference on Information and Knowledge Management, October 19--23, 2020, Galway, Ireland}
\acmPrice{15.00}
\acmISBN{978-1-4503-XXXX-X/18/06}

\begin{document}

\title{Knowledge Graph Validation}

\author{Elwin Huaman}
\email{elwin.huaman@sti2.at}
\affiliation{%
  \institution{Semantic Technology Institute}
  \city{Innsbruck}
  \state{Austria}
}

\author{Elias K\"arle}
\email{elias.kaerle@sti2.at}
\affiliation{%
  \institution{Semantic Technology Institute}
  \city{Innsbruck}
  \state{Austria}
}

\author{Dieter Fensel}
\email{dieter.fensel@sti2.at}
\affiliation{%
  \institution{Semantic Technology Institute}
  \city{Innsbruck}
  \state{Austria}
}

\renewcommand{\shortauthors}{E. Huaman et al.}

\begin{abstract}
Knowledge graphs (KGs) have shown to be an important asset of large companies like Google and Microsoft. KGs play an important role in providing structured and semantically rich information, making them available to people and machines, and supplying accurate, correct and reliable knowledge. To do so a critical task is knowledge validation, which measures whether statements from KGs are semantically correct and correspond to the so-called "real" world. In this paper, we provide an overview and review of the state-of-the-art approaches, methods and tools on knowledge validation for KGs, as well as an evaluation of them. As a result, we demonstrate a lack of reproducibility of tools results, give insights, and state our future research direction.
\end{abstract}





\maketitle

\section{Introduction}
\label{sec:introduction}
Knowledge curation (aka knowledge refinement)~\cite{FenselSAHKPTUW20} is the process of ensuring (or ideally improving) the quality of knowledge graphs (KGs). In this context, knowledge validation is a critical task to provide accurate, correct, and reliable knowledge.

The knowledge validation task in KGs is to decide whether an assertion (e.g., "Bill Gates is 64 years old") from a knowledge graph (KG) is semantically correct or not and whether it corresponds with the so-called "real" world. So far, validating KGs is often carried out by human experts (e.g. Wikidata), however, this approach cannot keep pace with the large size of KGs, which typically cover multiple domains and billions of facts~\cite{FenselSAHKPTUW20, HoganBCddGL2020, NoyGJNPT19}. There have been many approaches to develop specialised methods and tools, with the aim to semi-automatically validate facts (aka fact checking) in KGs. Recent knowledge validation studies can be categorized according to the validation target (e.g. instance assertions, property-value assertions and equality assertions) and can use the KG itself or an external knowledge source for validating the assertions \cite{Paulheim2017}. However, to the best of our knowledge, there is no comparative study of validation frameworks. Therefore, it is crucial and necessary to explore and evaluate efficient and effective semi-automatic methods and tools for assessing the reliability of facts in KGs, in other words, evaluating existing validation methods and tools. 

In this paper, we evaluate state-of-the-art validation frameworks aiming to find effective and efficient frameworks that can tackle the validation of KGs. We listed and described validation frameworks and evaluate the performance of selected ones. Furthermore, we state some remarks and discussion about the frameworks and future work.
In order to have a comprehensive survey of the current state-of-the-art validation frameworks, first, we describe methods, tools and approaches for knowledge validation, later on we define (1) research questions to perform our study, (2) criteria to reduce the number of frameworks to be evaluated, and (3) benchmark datasets, finally we setup and execute the selected frameworks. Last but not least, we discuss our results and findings by presenting an analysis of the frameworks, as well as answering our defined research questions.

The reminder of this paper is structured as follows: in Section 2, we present the literature review on knowledge validation methods, tools, and approaches. Section 3 describes our evaluation approach. The result and analysis of our evaluation is described in Section 4. Finally, we conclude in Section 5, providing remarks and future work.

\section{Literature Review}
\label{sec:literature-review}
In this section, we present the findings of the literature review regarding KG validation. We first define error sources based on a simple maximal knowledge representation formalism proposed in \cite{FenselSAHKPTUW20}. Afterwards, we present the review of approaches with respect to each error source. Last but not least, we describe validation methods and tools.

\subsection{Defining Error Sources}
\label{subsec:error-sources}
For producing a more consistent, accurate, and useful KG, the errors sources need to be identified and tackled, these error sources have been defined based on a maximal simple knowledge representation formalism in \cite{FenselSAHKPTUW20}. Which identify the following error sources:

\begin{itemize}
    \item \textbf{Instance assertions}: or $isElementOf(i,t)$ is wrong when the instance assertion is semantically wrong. Which can be handled by deleting assertion or finding proper $t$. The identification and correction of wrong instance assertions are complex tasks \cite{EstevesRRL2018, GangemiNPDMC2012, LiangXZHW2017, NuzzoleseGPC2012, PaulheimB2013, PaulheimB2014, SleemanF2013}.
    
    \item \textbf{Property value assertions}: or $p(i\textsubscript{1},i\textsubscript{2})$ is wrong when the property value assertion is semantically wrong. Which can be addressed by deleting or correcting assertion., i.e. define proper $i\textsubscript{2}$, or search for better $p$, or $i\textsubscript{1}$. The identification and correction of wrong property value assertions has been studied by several authors \cite{Debattista0A2016, Lertvittayakumjorn2017, MeloP2017}.
    
    \item \textbf{Equality assertions}: or $isSameAs(i\textsubscript{1},i\textsubscript{2})$ is wrong when the equality assertion is semantically wrong. That can be tackled by deleting or replacing assertion by a SKOS\footnote{\url{https://www.w3.org/TR/skos-reference/}} operator. Wrong equality assertion is not a straightforward task \cite{EstevesRRL2018, PernelleRS2018, RaadBHPS2018}.
\end{itemize}

We provide data consumers, data producers, and researchers with a literature summary of the existing work on KG validation with respect to each error source described above. Thereby, encouraging the development of new approaches, methods, or tools.

\subsection{Methods}
In this section, we survey methods for validating facts in KGs. We distinguish them according the error sources (i.e. instance assertion, property value assertion, and equality assertion).

\subsubsection{Methods for Validating Wrong Instance Assertions}
\label{subsec:wrong-instance}
There are several methods that can address the identification or correction of wrong instance assertions. For instance, methods that use association strength between subject and object of a triple \cite{JiaXCWE19}, discriminative path mining \cite{ShiW16, SyedRN2019}, information extraction \cite{SpeckN19}, topic coherence \cite{AletrasS13} techniques, as well as Wikipedia pages \cite{ErcanEH19}. 

\subsubsection{Methods for Validating Wrong Property Value Assertions}
\label{sec:wrong-property}
For identifying wrong property value assertions and correcting them there are methods that use knowledge stream \cite{ShiralkarFMC17}, outlier detection \cite{SyedRN2019, WienandP2014}, and property mining \cite{ShiW16} techniques. As well as, methods that use external sources like Wikipedia pages \cite{AletrasS13, ErcanEH19} and methods that are focused on numerical values \cite{ThorneV17, WienandP2014}.

\subsubsection{Methods for Validating Wrong Equality Assertions}
\label{sec:equality-instance}
To the best of our knowledge the literature review does not show approaches that tackle the validation of wrong equality assertions.

\begin{table*}
    \caption{Overview of validation frameworks}
      \label{tab:validation-frameworks}
    \begin{tabular}{lcccccccc}
        \toprule
        \textbf{Detail} & \textbf{COPAAL} & \textbf{DeFacto} & \textbf{ExFact} & \textbf{FactCheck} & \textbf{FacTify} & \textbf{Leopard} & \textbf{Surface} & \textbf{Tracy}\\
        \midrule
        Documentation & Yes	& Yes & No & Yes & No & No & No & No \\ \hline
  GUI & Yes & Yes & Yes & Yes & No & No & No & Yes \\ \hline
  Online & 
  \href{https://github.com/dice-group/COPAAL}{Github} &
  \href{https://github.com/DeFacto/DeFacto}{Github} &
  \href{https://www.dropbox.com/sh/wpyyiyy5lusph40/AAC72xbQoGhCu4Qpa-mwUvDua?dl=0}{Dropbox} &
  \href{https://github.com/dice-group/FactCheck}{Github} & 
  \href{http://qweb.cs.aau.dk/factify/}{File} &
  \href{https://github.com/dice-group/Leopard}{Github} &
  -- &
  \href{https://www.dropbox.com/sh/wpyyiyy5lusph40/AAC72xbQoGhCu4Qpa-mwUvDua?dl=0}{Dropbox} \\ \hline
  Programming language & Java & Java, Python & Java & Java & Python & Java & -- & Java \\ \hline
  Reference & \cite{SyedSRN19} & \cite{GerberELBUNS15} & \cite{GadElrab0UW2019} & \cite{SyedRN18} & \cite{ErcanEH19} & \cite{SpeckN19} & \cite{PadiaFF18} & \cite{GadElrab0UW2019b} \\ \hline
  Last update & 2019 &	2018 & 2018 & 2018 & 2019 & 2018 & 2018 & 2018 \\ 
      \bottomrule
    \end{tabular}
  \end{table*}
  
\subsection{Tools}
\label{subsec:tools}
From the survey in the previous section, we can observe that is different approaches proposed for KG validation. In the following, we give an overview of existing tools for validating assertions and some relevant approaches.

\subsubsection{COPAAL or Corroborative Fact Validation,}
validates facts relaying on statements that exist within a KG. Basically, (i) initialization, where COPAAL\footnote{\url{https://github.com/dice-group/COPAAL}}\cite{SyedRN2019} receives a triple, then (ii) path discovery, that identifies a set of alternate paths between the subject and object of the triple, (iii) path scoring, COPAAL scores each alternate path based on counting the number of paths which connect the subject and object of the triple, based on these scores it generates a final score to express the veracity of the given triple.

\subsubsection{DeFacto,}
or Deep Fact Validation\footnote{\url{https://github.com/DeFacto/DeFacto}} \cite{LehmannGMN12}, aims to validate triples by retrieving trustworthy sources from the web. It has two use cases: (1) find provenance information on the web, and (2) check the trustworthiness degree to which a triple is to be true. DeFacto works as follows, first, it finds webpages that can confirm a given triple, then, the trustworthiness confidence of these webpages are measured by calculating (a) Topic majority in the Web that represents the number of web pages having similar topics of a selected webpage, (b) Topic majority in search results that calculates a similarity of a given web page for all web pages based on a given triple, (c) Topic coverage that measures the ratio of all topic terms for a triple between all topic terms retrieved from a webpage, and (d) Pagerank\footnote{\url{https://en.wikipedia.org/wiki/PageRank}} that measures a relative importance of a webpage. In \cite{GerberELBUNS15}, DeFacto has implemented additional improvements such as temporal fact validation that estimates a timeframe in which a triple is or was valid, as well as, multilingual fact validation that finds evidence on resources written in English, German and French.

\subsubsection{FactCheck,}
uses Linked Open Data (LOD) sources to measure the degree to which a triple is to be true. First, FactCheck\footnote{\url{https://github.com/dice-group/FactCheck}} \cite{SyedRN18} takes as input an RDF triple, Second, it transforms the triple into "natural language" statement by bootstrapping the Web of Data\footnote{BOA extracts natural language patterns from The Web. \url{http://boa.aksw.org/}}, third, these patterns extracted are used for searching through a reference corpus and gathering documents containing similar statements. finally, a trustworthiness confidence value is calculated for each document. Besides, the trustworthiness confidence is calculated based on topic model technique.

\subsubsection{FacTify,}
aims to retrieve textual evidence for facts, FacTify\footnote{\url{http://qweb.cs.aau.dk/factify/}} \cite{ErcanEH19} combines exact matching and semantic matching. First, it transforms facts into a set of keywords, then, these keywords are used to retrieve documents that are ranked based on term frequency (i.e. they use OKAPI BM25\footnote{OKAPI BM25 ranks the relevance of documents, it is mostly used by search engines.}). Afterwards, it compares the semantic relatedness between them using Word embedding technique. furthermore, FacTify presents a benchmark for evaluation (See Section \ref{subsec:benchmark}).

\subsubsection{Leopard,}
is a framework that uses information extraction techniques to validate triples. Leopard\footnote{\url{https://github.com/dice-group/Leopard}} \cite{SpeckN19} starts the fact checking by (1) data acquisition, where a list of websites for crawling is provided and later on stored, (2) attribute extraction, this module extracts values from crawled pages for four attributes (e.g. phone number), then, it (3) scores and ranks using Fox \cite{SpeckN14a} (a named entity recognition framework).

\subsubsection{Surface,}
aims to verify facts by classifying them as supported, refuted, or unsure. Surface~\cite{PadiaFF18} follows three steps: given a fact (1) retrieves evidences (i.e. documents) from Wikipedia using string matching method; (2) determines utility of retrieved evidences using binary classification method; and (3) classifies facts using similarity metrics such as cosine and Jaccard similarities.

\subsubsection{S3K or Seeking Statement-Supporting top-K Witnesses,}
is a framework that retrieves documents in order to support a fact. Given a statement, S3K~\cite{MetzgerEHS11} retrieves documents that contain the given statement in different variations (e.g. synonyms or textual expressions about statements from a dictionary), and afterwards, it ranks matching documents according to statement frequency, pattern confidence, entity occurrences, and page authority.
    
\subsubsection{TISCO or Temporal Information SCoping,}
aims to determine the temporal scope of facts (i.e., the time intervals in which the fact is valid). TISCO\footnote{\url{http://tisco.disco.unimib.it/temporal-interval-scoping/}}~\cite{Rula19} follows three steps: (1) Temporal evidence extraction, extracts information for a given fact from the web and DBpedia. For achieving the extraction, it uses the DeFacto framework, which returns possible evidence for a given fact. Then it returns a list of all dates and their number of occurrences for a given fact. (2) Matching, applies a local and global approach for normalizing the time scope. Local normalization takes the relative frequency of a fact, and global normalization considers the frequency of all facts that share the same subject. The Matching function returns interval-to-fact significance matrix associated with a fact (i.e., a fact associated with several time intervals), and (3) Selection and Reasoning, select the time intervals associated with a fact. Once there is, a set of significance matrices, it applies two functions, which are Neighbour-x function that selects the neighborhood of the time interval with the maximum significance score and Top-k function that selects intervals whose significance is close enough to the most significant interval. Finally, TISCO uses Allen's interval algebra\footnote{Allen's interval algebra can define relations between time intervals.} to merge two-time intervals associated with a fact.

\subsubsection{More approaches:}
\label{subsec:more-approaches}
There are many ways in which KGs may be validated, depending not only on how high is the flow of knowledge between the subject and object of a triple in a KG but also how accurate is the triple concerning external knowledge sources.
\begin{itemize}
    \item Detecting incorrect numerical data in DBpedia \cite{WienandP2014}. The authors propose the use of unsupervised numerical outlier detection techniques which consists of two-steps, (1) group instances by their type and (2) apply outlier detection. They also applied a preprocesing strategy in cases where instances are missing or have wrongly assigned types, which consists of clustering by type vectors (i.e. use FeGeLOD framework\cite{PaulheimF12} that enriches resources with information gathered from LOD). Following a similar approach is presented by \cite{ThorneV17}.

    \item Evaluating topic Coherence Using distributional Semantics \cite{AletrasS13}, measures the degree to which a set of words generated by a topic model are coherent, this approach is used by \cite{SyedRN18}.
    
    \item Finding streams \cite{ShiralkarFMC17}, it takes a triple as input, it identifies the set of paths that produce "flow of knowledge" between the subject and object of the triple. For instance, each edge of the graph has a \textit{capacity} to carry knowledge between subject and object nodes of the triple and a \textit{cost} of usage a stream. The authors applies mining of the structure of the KG to define similarity between predicates, that it to say, the more similar, the higher flow of knowledge.
     
    \item KGTtm \cite{JiaXCWE19} (Knowledge Graph Triple trustworthiness measurement model) aims to generate a trustworthiness value for triples in a KG. It bases the trustworthiness degree on three aspects, (1) association strength between subject and object of the triple, (2) checking whether determined relationship (predicate) occur between the subject and object of a triple, it uses translation invariance method. (3) exploit reachable paths that can give credibility to the triple. A similar approach is applied by \cite{Xie0LL18} where they consider local triple confidence and global path confidence. 
    
    \item Property value mining \cite{ShiW16}, a path-based method that extracts, selects, and validates facts. First, it takes as input a triple, second, it extracts predicate paths that alternatively can connect the subject and objects of the triple, third, it selects the most discriminating predicate paths, and finally, it validates by comparing the input triple with triples that are built using alternative paths.
    
    \item ROXXI \cite{ElbassuoniHMS10} is a tool built in top of a knowledge base. ROXXI presents on its interface information of: a query module where users enter facts that need being validated, list of extracted documents based on facts, documents ranked based on occurrences of facts, generated snippets of information for each document, and a browser where users can read more about extracted facts.
    
    \item Trustworthiness of Web Search Results \cite{NakamuraKJOKTOT07}. The authors made a survey regarding the use of search engines with the aim of analysing trustworthiness factors of search results (i.e. the trustworthiness of a web page). Based on this analysis, they propose a prototype system, which calculates the trustworthiness degree of web search results. This system relies on (1) Topic Majority that evaluates the number of pages related to the query and the number of pages containing the same topic, (2) Topic coverage that calculates the number of topics a search result contains between the number of topics a query has, (3) Locality of link sources that analysis the geographic distribution of link sources, and (4) other information like publisher and last modified date.
\end{itemize}

A first observation is that these approaches can be classified into two categories 1) using the KG itself to create corroborative paths and 2) using external sources like Wikipedia, LOD, and search engines for validation of triples.

In addition, we also observe that most of the approaches focus on only one target i.e. property values, numerical values. Approaches that try to validate e.g. instance and property value are quite rare. Besides the fact that there exist no approaches that validate equality assertions.

An interesting finding is that, the reviewed approaches address KG validation and many of them are graph-based approaches. For instance, the input for many of the tools are transformed into a triple before to be validated.

\begin{figure*}
    \centering
    \includegraphics[width=1\linewidth]{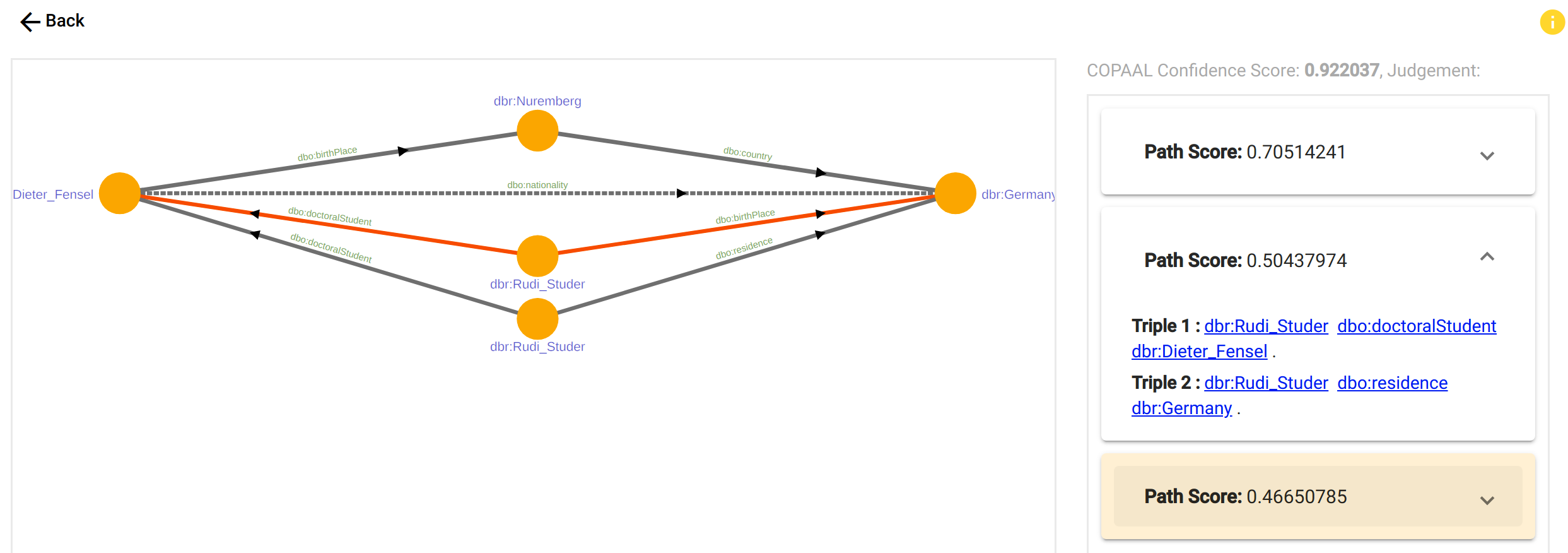}
    \caption{Screenshot of "COPAAL".}
    \label{fig:copaal}
\end{figure*}

\section{Evaluation Approach}
\label{sec:evaluation-approach}
In this section, we describe our evaluation approach, which starts by defining research questions in order to perform this study, afterwards we define criteria to select tools to be evaluated, later we identified benchmark datasets that can be used to evaluate the validation tools. Finally, we setup and execute the tools.

\subsection{Research Questions}
\label{subsec:research-questions}
In order to perform this study, the following research questions were formulated:

\begin{itemize}
    \item[RQ1:] \textbf{How does the framework validate statements?}
    A validation framework typically assigns a score of veracity to a given statement, the score typically ranges from $0$ to $1$, where a value of $0$ defines the minimum fulfilment degree of a statement regarding its veracity, a value of $1$ the maximum fulfilment degree.
    
    \item[RQ2:] \textbf{How flexible are the framework's configurations with respect to their features?}
    A feature of a validation framework can be represented as a module that allows to achieve some tasks. A validation framework for instance typically assume external KGs (e.g. DBpedia), dictionaries as reference sources.
    
    \item[RQ3:] \textbf{How scalable is the framework?}
    The scalability of the validation framework defines its applicability to large KGs, for instance, the effectiveness and efficiency of the validation frameworks.
\end{itemize}

\subsection{Tool Selection}
\label{subsec:tool-selection}
A set of validation frameworks (or tools) were found while we were doing the review of the literature (see Section \ref{sec:literature-review}). We list them in Table \ref{tab:validation-frameworks}.

Reproducibility is a very important part of evaluation \cite{BonattiDPP2018}. For instance, only if one can independently compare and verify the presented results of tools with the results obtained after trying the tool, one can say that it works and its results are verifiable. In other words, the validation frameworks must ensure their reproducibility.

Currently, reproducing validation framework results is challenging (almost impossible) because the frameworks were developed using different programming languages (e.g. Java, Python), they were only developed for: a competition (e.g. Leopard) or validating specific knowledge bases like DBpedia (e.g. COPAAL), they do not provide a manual or any documentation (e.g. ExFact and Factify), or they are just a prototype (e.g. Surface).

As we were assuming there is no single approach for validating KGs that can achieve high effectiveness and efficiency. Therefore, we collected a list of validation frameworks from the state-of-the-art (see Table \ref{tab:validation-frameworks}). We reduced the number of frameworks to be evaluated based on two criteria: (1) whether the framework is available online, this criterion is related to ease of access, which implies that the framework can be located, downloaded and updated, and (2) whether it has enough documentation to be executed, this criterion evaluates the availability of documentation, for instance, installation instructions, user guides, FAQs, and Wikis. In other words, the validation frameworks to be evaluated are COPAAL~\cite{SyedSRN19}, DeFacto~\cite{GerberELBUNS15}, and FactCheck~\cite{SyedRN18}.

\subsection{Benchmark Datasets}
\label{subsec:benchmark}

There are currently a limited number of published datasets resources for knowledge validation. \cite{VlachosR14} released 221 labeled claims in the political domain, \cite{ThorneVCM18} released a dataset containing 185K claims about properties of entities and concepts which were verified using articles from Wikipedia. Moreover,~\cite{HuynhP19} proposed a benchmark focused on data properties to evaluate the performance of validation algorithms.

\begin{itemize}
    \item FactBench\footnote{\url{https://github.com/DeFacto/FactBench}} (Fact Validation Benchmark), provides a multilingual (i.e. English, German and French) benchmark for evaluating fact validation algorithms. Facts provided by FactBench are scoped with a timespan in which they were true and describe several relations such as award, birth, death, foundation place, leader, publication date of books, and Spouse.
    
    \item Factify Benchmark\footnote{\url{http://qweb.cs.aau.dk/factify/}} contains 56 subgraphs (of 1.41 facts as average number of fact for each subgraph) extracted from YAGO knowledge base, there is 4,145 unique passages for these subgraphs. The domain of this benchmark is very generic, e.g., politics.
    
    \item FEVER\footnote{\url{https://github.com/sheffieldnlp/fever-naacl-2018}} \cite{ThorneVCM18} is a dataset that can support validation of facts. It has been extracted from Wikipedia and contains 185,445 claims manually verified and classified as supported, refuted, or notEnoughinfo. FEVER follows three steps: first retrieves relevant document, second sentence-level evidence selection by users, and third textual entailment based on term frequencies and TF-IDF cosine similarity between the claim and evidence.
    
\end{itemize}

For evaluation methodologies, our first observation is that these datasets are generated mainly during contests like FEVER\footnote{\url{http://fever.ai/}} (a Workshop on Fact Extraction and Verification) with specific purposes of measuring validation tools or approaches. Another interesting observation is that, Wikipedia is most frequently used by approaches described in section \ref{subsec:more-approaches} for validating facts \cite{AletrasS13, ErcanEH19, PadiaFF18}. As well as, only few of the approaches make retrospective evaluation for evaluating the resulted validation of approaches \cite{ElbassuoniHMS10}. Therefore, in order to make future works on KG validation comparable, it would be useful to have a common selection of benchmarks.

\subsection{Tool Setup and Execution}
\label{subsec:tool-setup}
The three selected frameworks, which are COPAAL, DeFacto and FactCheck, publish their source code on GitHub\footnote{\url{https://github.com/}} repositories. However, COPAAL and FactCheck have not released any version of their source code. Only DeFacto, which has provided three different versions along 2012, 2014, and 2015.
Furthermore, it has 5 different branches\footnote{A branch isolates development work without affecting other branches in the repository.}, which none of them is active\footnote{Active branches are branches that anyone has committed to within the last three months.}, moreover, its last source code update was made in September 2018. COPAAL and FactCheck have three and six non-active branches respectively.

The process of running the validation frameworks differ from each other, therefore, first we clone the frameworks' repositories, second we follow their usage instructions, and finally run the frameworks.

\begin{figure*}
  \centering
  \includegraphics[width=1\linewidth]{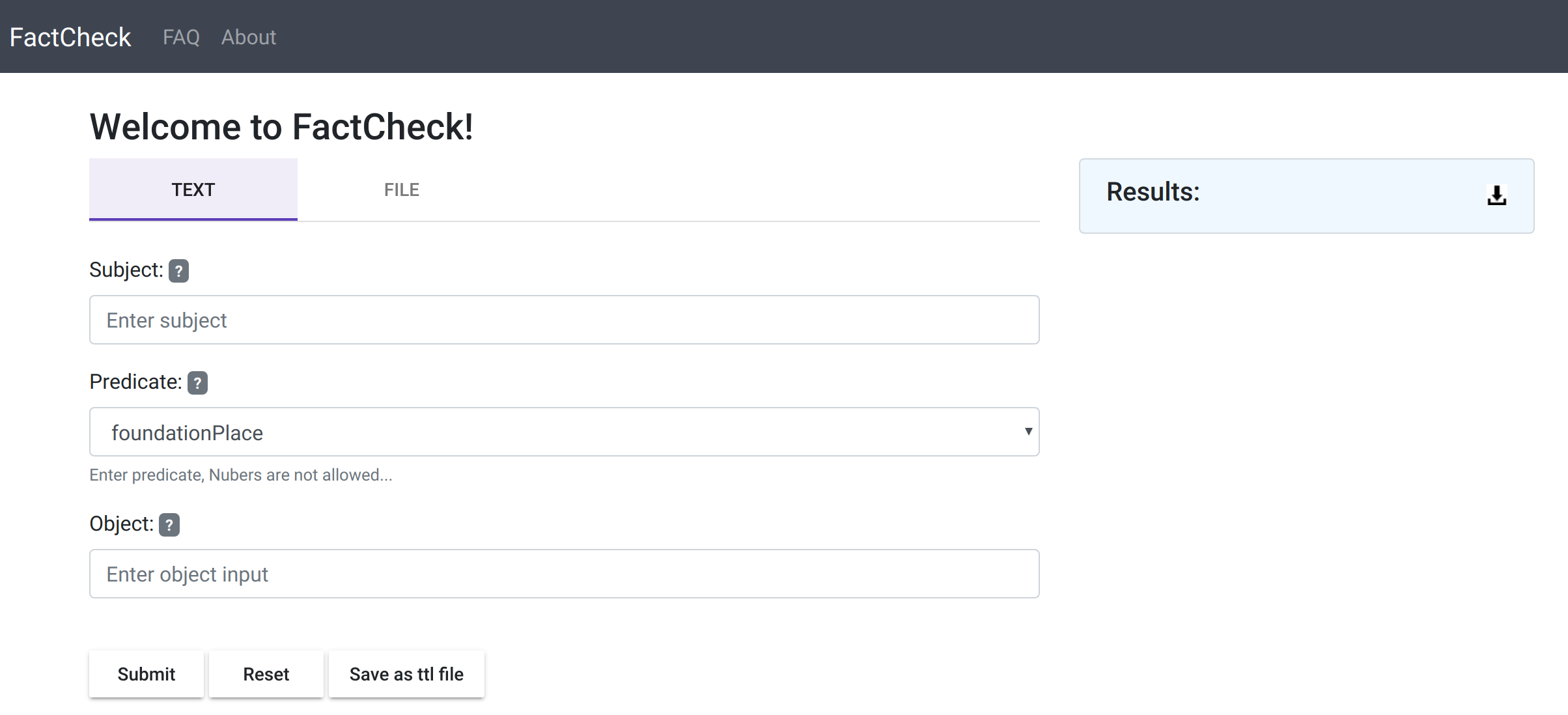}
  \caption{Screeshot of "FactCheck".}
  \label{fig:factcheck}
\end{figure*}

\section{Result}
\label{sec:result}
This section discusses the results obtained through this study, first we analyse the execution of frameworks, later we answer each research question defined in Section \ref{subsec:research-questions}.

\subsection{Analysis}
\label{subsec:analysis}

The COPAAL framework\footnote{\url{https://github.com/dice-group/COPAAL}} contains two components namely the service and the demo (UI-service). The service can be deployed using an Integrated Development Environment (IDE) as a packaged application or using the Maven package application, and the demo can deploy a web interface in order to query and validate facts. Figure \ref{fig:copaal} shows a screenshot of the COPAAL framework, for a given triple, for example, ("Dieter Fensel", "nationality", "Germany") COPAAL returns a confidence value of $0.922037$ that indicates the veracity\footnote{The veracity defines the reliability, correctness \cite{WangS1996} syntactically and semantically \cite{FarberBMR2018}, and truthfulness of the data~\cite{FenselSAHKPTUW20}.} of that triple. Moreover, the COPAAL framework shows a graph with a set of corroborative paths.

\begin{figure}
    \centering
    \includegraphics[width=1\linewidth]{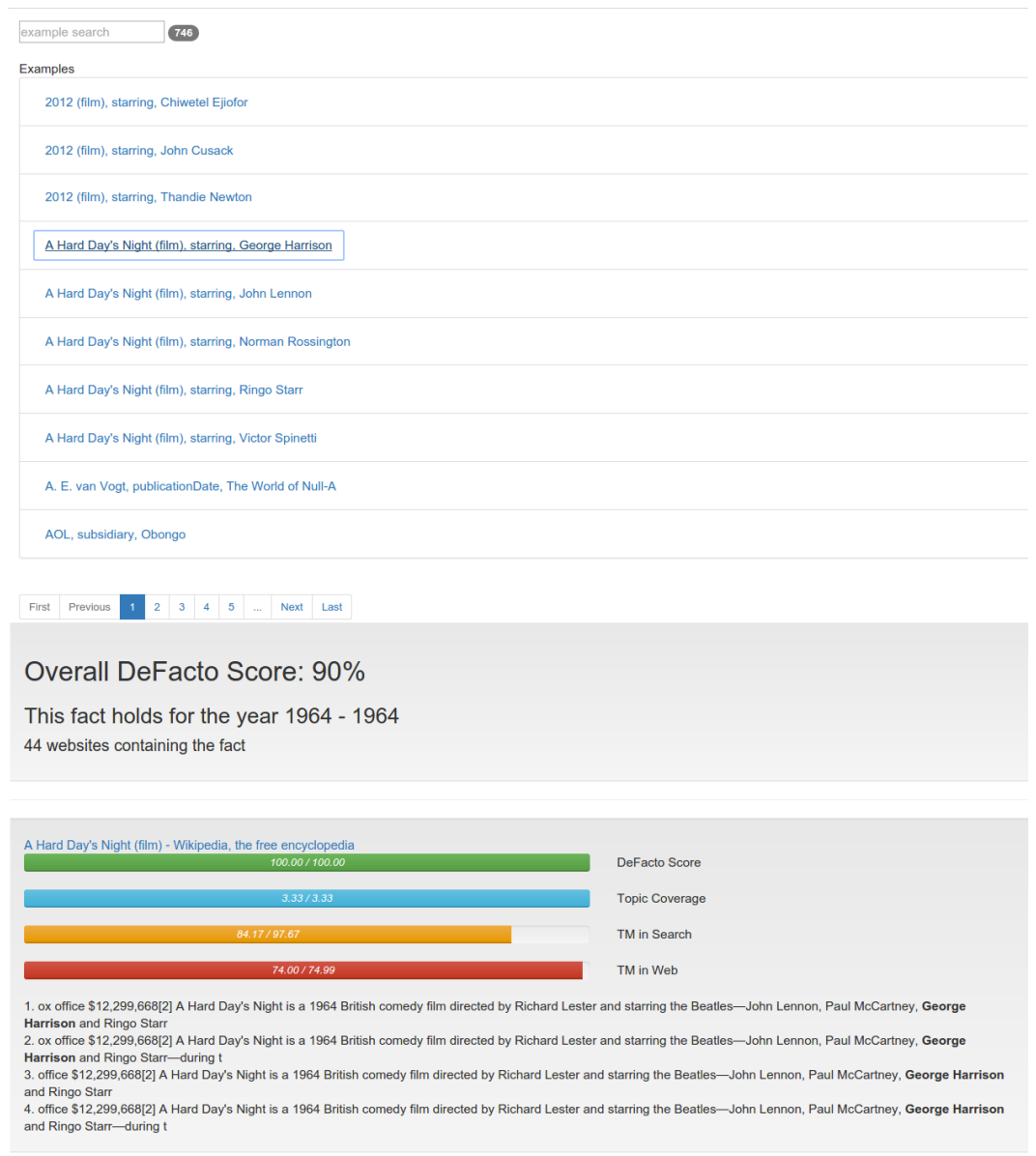}
    \caption{Screeshot of "DeFacto" \cite{GerberELBUNS15}.}
    \label{fig:defacto}
\end{figure}

The FactCheck and the DeFacto frameworks follow a similar approach and they provide the same installation process, moreover, they have three components that together make a validation framework. First, the framework-core that contains the core algorithm that is used to validate a given statement, second the framework-service which is an API for the framework which receives triples contained in the JSON request, and third the framework-demo (or front-end) which provides an HTML form to enter either a subject, predicate and object or upload a turtle\footnote{RDF 1.1 Turtle, see \url{https://www.w3.org/TR/turtle/}} file, and submit to the framework-service. Figures \ref{fig:defacto} and \ref{fig:factcheck} show screenshots of their interfaces.

After having cloned the frameworks' repositories and followed the usage instructions, we found the following remarkable issues:
\begin{itemize}
    \item The repositories do not provide a source code release during the last 4 years, which makes the executing of the framework very complicated. 
    \item The repositories have multiple branches and the documentation is not updated for any of them. For instance, they do not provide a Wiki\footnote{Wikis on GitHub help to present in-depth information about a project.} documentation about the branches.
    \item The framework's source code (i.e. master branch) in the repository does not provide documentation (e.g. COPAAL) or the documentation is referring to an old version (e.g. DeFacto and FactCheck).
    \item The frameworks are dependant on Google or Microsoft Bing services (e.g. DeFacto and FactCheck). Yet a local corpus (e.g. Wikipedia Dump\footnote{A Wikipedia dump provides a copy of all available content of Wikipedia.}) can be configured using an elasticsearch\footnote{\url{https://www.elastic.co/elasticsearch/}} instance.
\end{itemize}

Finally, after trying the validation frameworks and given the list of issues pointed out above, we were able to execute only the COPAAL validation framework. We defined a triple (i.e. ("Dieter Fensel", "nationality", "Germany")), we entered the triple to COPAAL and execute it. Afterwards COPAAL returned a graph that contains corroborative paths with a confidence score (see Figure \ref{fig:copaal}).

\subsection{Answering the Research Questions}
\label{sec:answering-rq}
In this section, we will answer the three research questions defined in Section \ref{subsec:research-questions}. Thus, we evaluate the frameworks w.r.t. their approach to validate facts, available features, and their performance.

\subsubsection{RQ1: How does the framework validate statements?}
COPAAL receives a triple (i.e. (subject, predicate, object) or simply (s,p,o)) and checks corroborative paths between the \textit{s} and \textit{o} of the triple in DBpedia, the object must be an instance. COOPAL evaluates the veracity of the given triple by computing the knowledge stream\footnote{"We call this set of paths a “stream” of knowledge" \cite{ShiralkarFMC17}.} \cite{ShiralkarFMC17}. The corroborative paths represent paths that have a high mutual information with the input triple. Moreover, COPAAL represents the output using graphs that are generated with the D3.js library\footnote{D3.js (\url{https://d3js.org/}) is a JavaScript library for producing dynamic and interactive data visualization.} and verbalized way of triples\footnote{"Verbalized RDF triples are sequences of sentences, which states the content of the corroborative paths in simple English sentences"~\cite{SyedRN2019}.}.

FactCheck receives a RDF triple, which is verbalized by means of using the rule-based LD2NL framework\footnote{LD2NL (\url{https://github.com/dice-group/ld2nl}) converts triples to natural language.} (used by COPAAL and DeFacto too), this verbalized sentences are used to search through a corpus (e.g. Wikipedia) that is indexed in a elasticsearch instance. Each document retrieved from the corpus is used as evidence for the given triple and used as input to generate a confidence value for the triple. FactCheck relies on the DeFacto tool in many of its implemented features \cite{SyedRN18}.

The DeFacto framework consists of five modules that allow users validating facts, given a triple or textual data as input: First, DeFacto retrieves the highest ranked web pages as evidence candidates, by querying search engines (e.g. Bing), the queries are verbalized using BOA framework\footnote{Bootstrapping approach for extracting RDF from unstructured data, see more \url{http://aksw.org/Projects/BOA.html}} \cite{GerberN12} (the same framework used by FactCheck); Second, DeFacto evaluates the retrieved web pages using information retrieval methods like string similarity; Third, DeFacto generates a confidence value that ranges between 0\% and 100\% where 0\% value means that there is no evidence for the given triple and 100\% that there is "much" evidences; Fourth, DeFacto enables exporting/storing the output as RDF; And fifth, DeFacto provides a front-end interface to perform the validation of facts (see Figure \ref{fig:factcheck}).

In summary, the frameworks' approaches have been described and they tackle the knowledge validation. Moreover, all the frameworks rely on corpora (either the Web and/or knowledge bases).

\subsubsection{RQ2: How flexible are the framework’s configurations with respect to their features?}

The validation frameworks present remarkable features. COPAAL has a module that verbalizes the corroborative paths output using the rule-based LD2NL framework, this framework converts triples to simple English sentences. FactCheck includes a \textit{Evidence classification} feature that classifies the retrieved documents with page title, total occurrence information. DeFacto implements a semi-automatic generation of search queries based on BOA framework. DeFacto has a module that supports the temporal validation of facts, this module relies on Wikipedia dumps as text corpora.

To sum up, the framework's features are complementary to each other, e.g., verbalize the triples, generate queries (e.g. queries with dates), later using them to search facts (e.g. temporal facts) in corpora, then classify retrieved results, and finally show the results by a GUI that might support graph visualization and verbalization of facts.

\subsubsection{RQ3: How scalable is the framework?}
The performance of COPAAL has been evaluated on a dataset, which contains information about the alma mater, birth place, death place, and educational degree of notable people \cite{ShiralkarFMC17}, the results shows that COPAAL performs poorly on not common used predicates like \textit{Death Place}. Moreover, the COPPAL verbalization module was evaluated by persons and linguists, the results shown that the fluency of the verbalized triples is still worthy of improvement.

FactCheck has generated a corpus of 420 million plain text documents (from Wikipedia and ClueWeb\footnote{\url{http://www.lemurproject.org/clueweb12/}}), which improves the F-measure of FactCheck and DeFacto on validating facts \cite{SyedRN18}.

In summary, the reproducibility of their results was not possible (see Section \ref{sec:conclusion}), instead of that, we analysed their results in the literature which indicate that improvements need to be done, e.g., expanding the corpus to obtain a better performance. Furthermore, the frameworks are not able to validate billions of facts, e.g. validating 274 facts took 13-15 minutes to COPAAL\footnote{See more: \url{https://github.com/dice-group/COPAAL}}.

\subsubsection{Summary}
The out-of-date state and the lack of documentation made the setup and execution of the frameworks almost impossible---which is also seen in other research areas such as entity resolution~\cite{HuamanKF2020}---, in addition to the fact that there is no official release of their source code. Therefore, after trying the validation frameworks, we were able to execute only the COPAAL validation framework. We can summarize the results as follows:

\begin{itemize}
    \item We try COPAAL by executing the examples provided in its repository. We upload a dataset called \textit{US\_Vice\_President}\footnote{The US\_Vice\_President dataset contains 274 facts, for example, (Barack\_Obama, vicePresident, William\_R.\_King), see more: \url{https://github.com/dice-group/COPAAL/blob/COPAAL-AFIRM/src/main/resources/US\_Vice\_President.nt}} into COPAAL to be validated, COPAAL relies on DBpedia's SPARQL endpoint to find corroborative paths that can support the accuracy of facts. After running COPAAL, it returned a 500 error that either means a time out error on the DBpedia's SPARQL endpoint or that the service fails/refuses to execute the query\footnote{See more: \url{https://www.w3.org/TR/sparql11-protocol/\#update-failure}}. This issue can be prevented by having a local repository of the knowledge source and execute the queries there.
    \item In a second attempted, we configure COPAAL with a different repository (Wikidata's SPARQL endpoint) and create two triples to be validated, namely ("Dieter Fensel", "country of citizenship", "United States") and ("Dieter Fensel", "country of citizenship", "Germany"). These triples were uploaded on COPAAL and it returns a truth value of $0.0$ on the two triples. It demonstrates that COPAAL needs a more complex configuration to deal with a different SPARQL endpoint.
    \item We executed the COPAAL-demo version that provides an interface where a user can fill in a form with a triple. We entered a triple e.g. ("Dieter Fensel", "nationality", "Germany") and got a graph of corroborative paths with a score (see Figure \ref{fig:copaal}). However, the interface only allows validating one triple at a time, which makes this approach not scalable for millions of triples.
\end{itemize}

The COPAAL framework has demonstrated that it can be easily setup and executed since it does not require a lot of dependencies. However, COPAAL cannot be easily adapted to different corpora or SPARQL endpoints other than DBpedia.



\section{Conclusion and Future Work}
\label{sec:conclusion}
In this paper, we have evaluated validation frameworks and compare their functionality, moreover whether they are able to tackle the validation of KGs. Through this study the following conclusions are achieved:

\begin{itemize}
    \item The frameworks have demonstrated not being easily reproducible (e.g. results of the frameworks can not be reproducible) or configurable to other corpora (e.g. COPAAL works only with DBpedia).
    \item The major shortcoming of validation frameworks is their dependence on proprietary services like Bing and Google search engines, leading to a higher cost of deployment. However, this disadvantage is common to mostly all validation frameworks. This can be prevented if a validation framework uses open corpora (e.g. Wikipedia) but its performance lows down (e.g. FactCheck).
    \item The validation frameworks present a set of features that can be fused somehow, for instance, a workflow of (1) verbalizing the triples to be validated, (2) semi-automatically generating queries (e.g. queries with dates), (3) using the generated queries to search facts (e.g. temporal facts) in corpora (either the Web and/or knowledge sources), (4) classifying retrieved results, and (5) showing the results by a GUI that might support graph visualization and verbalization of resulted facts.
\end{itemize}

We provided a literature review and evaluation of validation frameworks in the context of KGs, we believe that there is still work to do in this field and we encourage to the scientific community that we all together can increase the reproducibility of our research outcome. This way, we can assure the credibility of our research and the reuse of the invested effort. 

As a next step, we will develop a validation framework that implements all of the features found throughout this study as well as making the research outcome reproducible.

Although some benchmarks have been presented, there is no common benchmark to compare validation frameworks' both scalability (efficiency and efficacy) and reproducibility. Therefore, we will create a validation benchmark for the tourism domain, against which validation frameworks can be tested.

\bibliographystyle{ACM-Reference-Format}
\bibliography{main}

\end{document}